\documentclass[conference]{IEEEtran}
\IEEEoverridecommandlockouts
\usepackage{cite}
\usepackage{amsmath,amssymb,amsfonts,amsthm}
\usepackage{algorithm,algorithmic,listings}
\usepackage{xcolor}
\usepackage{xspace}
\usepackage{enumitem}
\usepackage{pifont}
\usepackage{multirow}
\usepackage{hyperref}
\usepackage{varioref}

\usepackage{graphicx}
\usepackage{textcomp}
\usepackage{flushend}

\usepackage{tikz,tkz-euclide,pdftexcmds,pgfplots}
\usetikzlibrary{backgrounds, tikzmark}
\pgfdeclarelayer{foreground}
\pgfsetlayers{background,main,foreground}
\usetkzobj{all}

\usepackage{tcolorbox}
\definecolor{mycolor}{rgb}{0.122, 0.435, 0.698}
\definecolor{gray1}{gray}{0.3}

\definecolor{darkgreen}{rgb}{0.0, 0.5, 0.0}
\definecolor{darkred}{rgb}{0.82, 0.1, 0.26}
\newcommand{\cmark}{\textcolor{darkgreen}{\ding{51}}}%
\newcommand{\xmark}{\textcolor{darkred}{\ding{55}}}%

\newcommand{\result}[1]{%
\begin{tcolorbox}[colframe=mycolor,boxrule=0.5pt,arc=4pt,
      left=6pt,right=6pt,top=6pt,bottom=6pt,boxsep=0pt,width=\columnwidth]%
      {#1}
\end{tcolorbox}%
}

\usepackage{calc}
\newlength\listingnumberwidth
\newcommand{\todo}[1]{}
\renewcommand{\todo}[1]{{\color{red} TODO: {#1}}}

\newcommand{\tool}{\textsc{Learn2fix}\xspace}

\def\BibTeX{{\rm B\kern-.05em{\sc i\kern-.025em b}\kern-.08em
    T\kern-.1667em\lower.7ex\hbox{E}\kern-.125emX}}
\begin{document}

\title{Human-In-The-Loop Automatic Program Repair\vspace{-0.25cm}}

\author{\IEEEauthorblockN{Marcel B\"{o}hme\quad\quad Charaka Geethal \quad\quad Van-Thuan Pham}
\IEEEauthorblockA{Monash University, Australia\\{\{marcel.boehme, charaka.kapugamawasangamagedon, thuan.pham\}@monash.edu}}
}

\maketitle

\begin{abstract}
We introduce \textsc{Learn2fix}, the first human-in-the-loop, semi-automatic repair technique when no bug oracle--except for the user who is reporting the bug--is available.
Our approach negotiates with the user the condition under which the bug is observed. Only when a budget of queries to the user is exhausted, it attempts to repair the bug. A \emph{query} can be thought of as the following question: ``When executing this alternative test input, the program produces the following output; is the bug observed''?
Through systematic queries, \textsc{Learn2fix} trains an \emph{automatic bug oracle} that becomes increasingly more accurate in predicting the user's response. Our key challenge is to maximize the oracle's accuracy in predicting which tests are bug-revealing given a small budget of queries.  
From the alternative tests that were labeled by the user, test-driven automatic repair produces the patch.

Our experiments demonstrate that \textsc{Learn2fix} learns a sufficiently accurate automatic oracle with a reasonably low labeling effort (lt. 20 queries). Given \textsc{Learn2fix}'s test suite, the GenProg test-driven repair tool produces a higher-quality patch (i.e., passing a larger proportion of validation tests) than using manual test suites provided with the repair benchmark.
\end{abstract}

\section{Introduction}
Automatic program repair (APR) \cite{survey1,survey2} holds the promise of automating the tedious, manual task of patching bugs. In their seminal paper, Le~Goues and colleagues \cite{genprog} demonstrated that APR is both feasible and cost-effective even at the scale of several million lines of code. Given a failing test suite, APR changes the buggy program such that all test cases pass. 

However, what if no such test suite is available? Suppose, a user reports a bug and provides a test input to reproduce the bug. We envision a semi-automatic approach that keeps the human-in-the-loop and negotiates the condition under which the bug is observed before repairing the bug. Strategically, the user is asked: ``\emph{For this other input, the program produces that output; is the bug observed}''? While the user might not have the expertise to understand the source code or to produce a patch, it seems reasonable to ask to distinguish expected from unexpected program behavior. Iteratively, an \emph{automatic bug oracle} is trained to predict the user's responses with increasing accuracy. Using the trained oracle, the user can be asked more strategically. Our key challenge is to maximize the oracle's accuracy, given only a limited number of queries to the user. 

In this paper, we introduce \tool, a technique that realizes our approach for programs that take numeric inputs. \tool uses \emph{mutational fuzzing} to generate alternative test inputs from the failing test, \emph{active learning} to construct a Satisfiability Modulo Linear Real Arithmetic SMT(LRA) constraint that is satisfied only by failing test inputs, and \emph{test-driven program repair} to fix the bug using the labeled tests. The hope is that test cases---sufficient to train a machine to accurately identify the bug---are sufficient to repair the bug.  

\textbf{How to generate more failing test cases}?
Starting from one failing test, our first objective is to increase the evidence for the bug before attempting to fix it. Metaphorically speaking, by exploring the ``neighborhood'' of the original failing test, we can stake out the ``boundaries'' of the bug and find more failing tests in the ``vicinity''. \tool uses mutational fuzzing to generate new tests from the failing one. In mutational fuzzing, a test input is modelled as a sequence of numbers (i.e., bytes or integers), and new test inputs are generated by applying mutation operators at random positions in the sequence.

\textbf{How do we know whether a generated test reveals the bug} (i.e., whether its a failing test case)?
At first, \tool asks the user to classify. However, it would be impractical to ask the human oracle for every generated test input whether the program produces the expected output. Instead, \tool \emph{trains an automatic oracle} that is asked instead. That is, it \emph{learns an SMT constraint} \cite{incal} that is satisfied only by failing test cases. Satisfiability Modulo Theory (SMT) provides a natural representation of program semantics and is a fundamental building block of symbolic execution \cite{dart,klee,jpf} and semantic program repair \cite{semfix,angelix}. 

\textbf{How do we train the automatic oracle most efficiently}? We suggest to maximize the probability that the user is asked to classify only test cases that are actually failing. Firstly, bugs are (hopefully) rarely observed, and hence, most generated test cases are actually passing. This yields a \emph{class imbalance problem}: The class of passing test cases is often substantially larger. Given insufficient evidence for the bug (i.e., a relatively small number of failing test cases), the trained oracle may exhibit a high error rate when classifying failing test cases. Secondly, we cannot ask the user for every test case whether it is failing. In both cases, the most reasonable strategy would be to present the user only with test cases that have a high probability of being labeled as failing.

\textbf{How do we know the probability that a test case fails before we can even ask the user}?
An automatic oracle only provides hard, binary decisions: Either the test case passes \emph{or} it fails. So, how can we derive the \emph{probability} that test $t$ fails? \textsc{Learn2fix} constructs an \emph{unbiased committee of oracles} from the original oracle, asks each member of that committee about the label of $t$, and estimates the probability that $t$ is failing as the proportion of members that classify $t$ as failing. If that probability is greater than one half, then $t$ is presented to the user. This approach is inspired by a seminal paper on reducing the labeling cost in image classification \cite{Holub2008}.\footnote{We present a \emph{randomized} variant of the algorithm presented by Holub et al. \cite{Holub2008}. We do not need to construct two classifiers for \emph{each} unlabeled data point in a fixed-size sample. Instead, our sample is continuously generated.}

\vspace{0.05cm}
\noindent
Our \emph{experiments} with Codeflaws \cite{codeflaws} demonstrate that\vspace{-0.04cm}
\begin{itemize}[itemsep=0.1cm]
    \item \textbf{Oracle quality is high}. Even though \tool has only ever seen a single failing test case from the manually labeled validation test suite, the automatic oracle is able to accurately predict the label of 75--84\% validation tests (and 76-80\% of \emph{failing} validation tests) for the median subject. The prediction accuracy further increases as more queries can be sent to the human oracle. 
    
  \item \textbf{The labeling effort is low}. As oracle quality increases, the proportion of generated tests that are sent to the human oracle decreases. This suggests that the automatic oracle gradually takes over from the human oracle. Meanwhile, the probability that a test case which is sent to the human is \emph{failing} increases. This suggests, our auto\-matic oracle is effective in reducing the number of queries.
  
  \item \textbf{The repair quality is high}. We used the GenProg automatic repair tool \cite{genprog} to (i)~repair each buggy program with the \emph{manually} constructed test suite (excl. \texttt{heldout.*}), and (ii)~repair each buggy program with the \emph{automatically} constructed test suite that is produced by \tool. While \tool produces fewer repairs, patch quality is higher. The proportion of validation tests that fail on the repair is 31\% smaller, on the average, than for GenProg when started with the manually constructed test suites. 
\end{itemize}

\noindent
In summary, our paper makes the following \emph{contributions}:
\begin{itemize}[itemsep=2pt]
  \item \textbf{Active Oracle Learning}. We introduce an active learning approach that infers an automatic oracle from well-placed labeling queries to a human oracle. The automatic oracle is represented as SMT constraint which is satisfied by all test inputs that have been labeled as failing. We address the \emph{class imbalance problem} by prioritizing the minority class of failing tests. Our experiments demonstrate that the inferred oracle accurately classifies a large proportion of (failing) validation tests.
  \item \textbf{Semi-Automatic Repair}. We develop the first human-in-the-loop program repair technique which negotiates with the reporting user the conditions under which a functional bug is observed before attempting to repair the program. Our experiments show that \tool's patches have a quality that is high while human effort is relatively low.
  \item \textbf{Overfitting}. Patches may be plausible but incorrect \cite{plausible,overfitting}. In our case, the patch would overfit to the failing test case. To address overfitting, we propose to co-evolve an automatic oracle and a labeled test suite. The generated labeled test suite augments the provided test suite.
  \item \textbf{Evaluation}. We implemented and evaluated \tool. To facilitate reproducibility, we make the implementation, all data, and scripts available.  
\end{itemize}

\section{Motivating Example}\label{sec:motivating}
We introduce the existing challenges of automatic program repair using a motivating example that is shown in \autoref{lst:triangle}. The example is taken from an experiment by Russ Williams \cite{russcon} who asked 12 participants to submit, together with some inputs and expected outputs, a solution to this problem:
\result{Implement a function \texttt{classify} which takes 3 inputs that represent the lengths of the sides of a triangle and returns an integer where the return value\vspace{-0.1cm}
\begin{itemize}
  \item 1 means it is equilateral (all sides equal length)
  \item 2 means it is isosceles (exactly 2 equal sides)
  \item 3 means it is scalene (no equal sides)
  \item 4 means it is an illegal triangle 
\end{itemize}\vspace{-0.1cm}
} 

\emph{Functional bug}. The participants submitted 22 programs out of which only 4 appear to be correct and 13 test suites totaling 5636 test cases. \autoref{lst:triangle} shows the submission by Steve who wrote \texttt{a==b==c} instead of \texttt{a==b\&\&b==c} in Line 6. For instance, given input $t=\langle 2,2,2\rangle$, Line~6 evaluates to \texttt{(2==2)==2} which evaluates to \texttt{(1)==2} and finally to \texttt{0}. Hence, Steve's program is incorrect for all equilateral triangles, except $\langle 1,1,1\rangle$, and for all isosceles triangles where \texttt{c==1}. For test input $t$, \texttt{classify} returns 2 (isosceles) while we expect it to return 1 (equilateral). Because of this difference between actual and expected output, we call $t$ as \emph{failing test case}. It is a witness of Steve's bug in Steve's program.
\lstset{%
   		framexleftmargin=5mm,
   		frame=single,
   		numbers=left,
   		numberstyle=\tiny,
   		basicstyle=\ttfamily\selectfont\small,
   	  frameround= tttt,
   	  breaklines = true,
   	  tabsize=2,
   	  captionpos=b,
   	  float,
   	  language=C}
\begin{lstlisting}[frame=single, caption={Buggy triangle classification program. Given the lengths\\
of all three sides, \texttt{classify} returns 1 if the triangle is equilateral,\\
2 if it is isosceles, 3 if it is scalene, and 4 if the triangle is invalid.\\ 
This program classifies almost all equilateral triangles as isosceles.\\ 
For example, \texttt{classify(2,2,2)} returns 2 while we expect 1.}, label=lst:triangle]
int classify(int a, int b, int c) {
  if (a <= 0 || b <= 0 || c <= 0)
    return 4;               // invalid
  if (a <= c-b || b <= a-c || c <= b-a)
    return 4;               // invalid
  if (a == b == c)          // BUG!
    return 1;               // equilateral
  if (a == b || b == c || c == a)
    return 2;               // isosceles
  return 3;                 // scalene
}
\end{lstlisting}

\emph{Automatic oracle}. Steve's program fails for all inputs that satisfy the following linear arithmetic constraint
\begin{align}
 [(a = b) &\wedge (b = c) \wedge  (a \neq 1) \wedge (o = 2)]\nonumber\\
 \vee [(a = b) &\wedge (c = 1) \wedge (a \neq 1) \wedge (o = 1)]\label{eqn:oracle}
\end{align}
where $o=\text{\texttt{classify}}(a,b,c)$ is the actual output. We call this an \emph{automatic oracle} for Steve's bug because it identifies for all inputs whether the Steve's bug is exposed. 

\emph{Automatic repair}. Given a sufficient number of test cases, an automatic repair tool such as GenProg \cite{genprog} or Semfix \cite{semfix} would first \emph{localize} Line~6 as fix location. Most failing and least passing test cases actually execute that statement. The repair tool would then \emph{repair} Line~6 such that all test cases are passing. However, we assume that there exists only a single failing test case. Even if Line~6 was identified as fix location, the produced patch may be plausible but incorrect \cite{overfitting}. Substituting the if-statement in Line~6 with \texttt{if(a==2)} would turn the test case $\langle 2,2,2\rangle$ into a passing one. However, the patch is overfitting and actually introduces a different bug.

\emph{Oracle problem}. If an automatic oracle was available, more failing test cases could be generated to augment the existing test suite \cite{fix2fit}. However, for functional bugs, such as the one in \autoref{lst:triangle}, the user that is reporting the bug or the developers are the only oracles  available to distinguish expected from unexpected behaviors. In this paper, we will discuss an active learning approach to automatically derive an automatic oracle, similar to the one in Equation~(\ref{eqn:oracle}).

\section{Learning to Decide Test Case Failure}
Given a buggy program, a failing test case, and only the user or developer as a test oracle, our technique \tool generates an automatic oracle and a set of labeled test cases. A \emph{test oracle} decides whether or not the  program produces the expected output for a given test input. In order to minimize the queries to the human oracle, \tool aims to maximize the probability that the human is presented with a failing test.

\setlength{\textfloatsep}{1\baselineskip}
\begin{algorithm}[t]
\renewcommand{\algorithmicrequire}{\textbf{Input:}}
\renewcommand{\algorithmicensure}{\textbf{return}}
\caption{Active Oracle Learning}\label{alg:overview}
\begin{algorithmic}[1] 
\REQUIRE Buggy program $\mathcal{P}$, Failing test case $t_\text{\xmark}=\langle \vec{i}, o\rangle$
\REQUIRE Human oracle $\mathcal{H}$, Maximimum labeling effort $l$
\STATE Failing test  $T_\text{\xmark}=\{t_\text{\xmark}\}$
\STATE Labeled tests  $T=\{t_\text{\xmark}\}$
\STATE Automatic oracle $\mathcal{O}=\text{\textsc{smt\_learn}}(T)$
\WHILE{$(|T| < l)$ and not timed out}
  \STATE Failing test  $t'_\text{\xmark} = \text{\textsc{select}}(T_\text{\xmark})$
  \STATE Generated test  $t_\text{\textbf{?}} = \text{\textsc{fuzz}}(t'_\text{\xmark})$
  \IF{$\text{\textsc{decide2label}}(t_\text{\textbf{?}}, \mathcal{O})$}
    \STATE Labeled test  $t=\mathcal{H}(t_\text{\textbf{?}})$
    \STATE Labeled tests $T=T\cup \{t\}$
    \STATE Automatic oracle $\mathcal{O}=\text{\textsc{smt\_learn}}(T)$
    \IF{$t$ is labeled as failing}
      \STATE Failing tests $T_\text{\xmark}=T_\text{\xmark}\cup \{t\}$
    \ENDIF
  \ENDIF
\ENDWHILE
\ENSURE Labeled test cases $T$, Automatic test oracle $\mathcal{O}$
\end{algorithmic}
\end{algorithm}

An overview of \tool is shown in Algorithm~\ref{alg:overview}. As input, \tool takes the buggy program $\mathcal{P}$, the failing test case $t_\text{\xmark}=\langle \vec{i}, o\rangle$ where $\vec{i}$ is a vector of input variable values and $o=\mathcal{P}(\vec{i})$ is the \emph{actual} output of $\mathcal{P}$ for $\vec{i}$. Next, we assume that there exists a human test oracle $\mathcal{H}$ that can decide whether a generated test is failing, i.e., exposes the bug. We also assume that $\mathcal{H}$ accepts at most $l$ queries. As output, \tool produces a set of labeled tests $T$ and an automatic test oracle $\mathcal{O}$ that can decide whether a test is failing without human intervention.

Algorithm~\ref{alg:overview} maintains two sets of test cases $T$ and $T_\text{\xmark}$ that are labeled overall and are labeled as failing, respectively. A \emph{test case} is a tuple consisting of (i)~a vector representing input variable values and (ii)~the program's actual output for that input vector (e.g., $t_\text{\xmark}=\langle \vec{i}, o\rangle$). The first automatic oracle $\mathcal{O}$ is learned by applying \textsc{smt\_learn} to the set of labeled test cases (Line~3; \autoref{sec:smtoracle}). Given only a single failing test case, at first, $\mathcal{O}=true$ identifies all test cases as failing. 

Algorithm~\ref{alg:overview} generates new test inputs, asks $\mathcal{H}$ to label certain generated test inputs, and refines the automatic oracle until the maximal allowable number of labels $l$ is reached or a timeout occurs (Lines 4--15). In each iteration, a new test input $t_\text{?}$ is generated from a failing one $t'_\text{\xmark}\in T_\text{\xmark}$ (Lines~5--6; \autoref{sec:fuzzing}). If \textsc{decide2label} decides that $t_\text{?}$ should be labeled, then the human oracle $\mathcal{H}$ is asked for the label, the labeled test is added to $T$ (and $T_\text{\xmark}$), and the automatic oracle $\mathcal{O}$ is retrained (Lines 7--14; \autoref{sec:learning}).

\subsection{Mutational Fuzzing to Generate More Failing Test Cases}\label{sec:fuzzing}
More failing test cases are needed. If the automatic oracle $\mathcal{O}$ is trained only with one failing test, the simplest oracle is $\mathcal{O}=\emph{true}$: All test cases fail, including the provided one. Of course, this oracle is highly inaccurate. In order to improve the accuracy of our automatic oracle, \tool must generate more test cases that---when labeled---become positive and negative evidence of the bug.

In order to generate more test cases in the ``vicinity'' of a failing test case, \tool employs mutational fuzzing. A \emph{mutational fuzzer} generates new test inputs by applying a set of mutation operators to an existing seed input. In our case, \textbf{\textsc{select}}$(T_\text{\xmark})$ in Algorithm~\ref{alg:overview} first selects as seed input a random test case $t'_\text{\xmark}\in T_\text{\xmark}$ while \textbf{\textsc{fuzz}}$(t'_\text{\xmark})$ applies mutation operators to $t'_\text{\xmark}$ such as bit flips, simple arithmetics, boundary values, and block deletion and insertion strategies to generate a new input $t_\text{\textbf{?}}$.\footnote{\scriptsize\url{https://lcamtuf.blogspot.com/2014/08/binary-fuzzing-strategies-what-works.html}}
Our implementation of \tool (i)~assumes that the input vector $\vec{i}$ has a fixed length, i.e., all generated test inputs have the same length,\footnote{Otherwise, we cannot fix the value domain during oracle-learning \cite{incal}.} and (ii)~for each position $a$ in $\vec{i}$, \tool either maintains the same value, adds or multiplies a constant or random value for $\vec{i}[a]$ to generate a new input $\vec{i}'$, such that $t_\text{\textbf{?}}=\langle\vec{i}',o'\rangle$ where $o'=\mathcal{P}(\vec{i}')$.  

The idea of exploring the neighborhood of a failing test case in order to collect more evidence of the location and behavior of the observed bug is not new. In fact, the widely successful, coverage-based, mutational greybox fuzzer American Fuzzy Lop (AFL) features a \emph{crash exploration mode}\footnote{\scriptsize\url{https://lcamtuf.blogspot.com/2014/11/afl-fuzz-crash-exploration-mode.html}} which allows to generate more crashing inputs from a seed crashing input. 

\vspace{0.1cm}
\textbf{Example}. For our motivating example in \autoref{lst:triangle}, starting with the failing test case $t_\text{\xmark}=\langle\langle 2,2,2\rangle,2\rangle$, for illustration suppose for each position $a$ in $\vec{i}$, we employ one of three mutation operators chosen uniformely at random: $\vec{i}'[a]=\vec{i}[a]$, $\vec{i}'[a]=\vec{i}[a]+1$, or $\vec{i}'[a]=\vec{i}[a]-1$. The following test cases $t_0,..,t_9$ are generated when actually running the fuzzer on $t_\text{\xmark}$:
%
{\small%
\begin{align}
\langle\langle 2,2,1\rangle,1\rangle_\text{\textbf{?}},&\quad
\langle\langle 1,3,3\rangle,2\rangle_\text{\textbf{?}},\nonumber\\
\langle\langle 1,3,2\rangle,4\rangle_\text{\textbf{?}},&\quad
\langle\langle 3,3,1\rangle,1\rangle_\text{\textbf{?}},\nonumber\\
\langle\langle 2,1,3\rangle,4\rangle_\text{\textbf{?}},&\quad
\langle\langle 3,3,3\rangle,2\rangle_\text{\textbf{?}},\nonumber\\
\langle\langle 2,1,1\rangle,4\rangle_\text{\textbf{?}},&\quad
\langle\langle 1,2,3\rangle,4\rangle_\text{\textbf{?}},\nonumber\\
\langle\langle 3,2,2\rangle,2\rangle_\text{\textbf{?}},&\quad
\langle\langle 2,3,2\rangle,2\rangle_\text{\textbf{?}}\nonumber
\end{align}}\vspace{-0.3cm}

The probability to generate another \emph{failing} test case by mutational fuzzing is much higher than by generational fuzzing. Suppose, we would randomly generate three numbers in the range $[-2^{63},2^{63}-1]$. The probability that the corresponding test case represents an isosceles triangle with $c=1$ or an equilateral triangle required to expose Steve's bug is infinitesimal. On the other hand, three of the ten test cases generated by our mutational fuzzer expose Steve's bug (if labeled by $\mathcal{H}$), i.e., $\langle\langle 2,2,1\rangle,1\rangle$, $\langle\langle 3,3,1\rangle,1\rangle$, and $\langle\langle 3,3,3\rangle,2\rangle$.
As more test cases are generated and labeled by the human oracle, the quality of the automatic oracle improves.

\subsection{Active SMT-based Oracle Learning}\label{sec:smtoracle}
An automatic oracle is needed. We cannot expect the user $\mathcal{H}$ to label every generated test case. Instead, \tool trains an automatic oracle $\mathcal{O}$ based on the tests that have already been labeled by $\mathcal{H}$. Like the human oracle, the \emph{automatic oracle} decides whether a given test is labeled as passing or failing. As the accuracy of the automatic oracle improves, $\mathcal{O}$ gradually takes over from $\mathcal{H}$. The automatic oracle is trained within an \emph{active learning loop} with the human oracle as the teacher and the automatic oracle as the learner.

\textbf{Why SMT}? We believe that Satisfiability Modulo Theory is a natural representation of buggy program semantics.
Firstly, SMT constraints are fundamental building blocks in semantic analysis, including symbolic execution and semantic program repair. \emph{Symbolic execution} \cite{dart,klee,jpf} uses SMT constraints (i.e., path conditions) to group all inputs that exercise the same path. The negation of the constituent branch conditions and the solution of the resulting constraints facilitates the exploration of alternative paths. \emph{Semantic program repair} \cite{semfix,angelix} uses SMT constraints for statement-level specification inference at the fix location and SMT-based synthesis to satisfy the inferred specification \cite{synthesis}. Unlike other binary classifiers, such as Support Vector Machines (SVMs), SMT constraints nicely capture the ``discontinuous'' nature of program behavior.\footnote{Our preliminary experiments with popular classifiers (e.g., SVM) revealed a prediction accuracy well above 90\%. However, upon closer inspection, we found that the classifier would conservatively predict \emph{all} test cases as passing. Even from the training set, almost none of the failing test cases was correctly predicted as failing. Indeed, in the presence of the \emph{class imbalance problem}, prediction accuracy is an improper measure of classifier quality \cite{classimbalance}. An SMT constraint, in contrast to regression-based classifiers, interpolates the training set, i.e., the constraint is satisfied by \emph{all} failing test cases and \emph{no} passing test case in the training set. In our results, we report the proportion of correctly identified \emph{failing test cases} in a validation set as measure of classifier quality.}

Secondly, our automatic oracle is a simple constraint. It only reflects the circumstances under which the program behaves in an \emph{unexpected} manner. It does \emph{not} reflect how the program is expected to behave. Indeed, this would require a full-fledged specification inference. For our motivating example in \autoref{lst:triangle}, we know that Steve's program provides unexpected results for all equilateral triangles, except for those with sides of unit length, and for all isosceles triangles where the third side has unit length. As SMT constraint, the automatic oracle $\mathcal{O}$ for Steve's bug is shown in Equation~(\ref{eqn:oracle}). It is not necessary for $\mathcal{O}$ to capture what we expect, e.g., to be a scalene triangle.

\textbf{Oracle inferencing}. \tool uses \textsc{Incal} to implement \textsc{smt\_learn} in Algorithm~\autoref{alg:overview}. \textsc{Incal}  \cite{incal} is a passive machine learning technique that given positive and negative examples learns an SMT(LRA) constraint that is satisfied by all positive but not by any negative examples. \emph{Satisfiability Modulo Linear Real Arithmetic} SMT(LRA) is a formalism that combines pro\-po\-si\-ti\-onal logic with expressions from linear arithmetic (e.g., [in]equalities, sums, and products) over continuous variables.

\textsc{Incal} casts the problem of generating an SMT constraint---that would satisfy all positive and no negative examples---itself as a \emph{satisfiability problem} (rather than an optimization problem). The authors argue that ``searching for a satisfying formula is usually faster than searching for an optimal one in practice''. It is assumed that all examples are \emph{labeled correctly}. It is guaranteed that \textsc{Incal} ``will find a constraint that satisfies all positive examples and none of the negatives, iff such a formula exists''.
In contrast to most rule learning approaches, \textsc{Incal} is \emph{non-greedy}, i.e., the formula is learned in one step rather than piece-by-piece.

\tool extends \textsc{Incal}s passive learning into an \emph{active learning} approach. The learner (i.e., automatic oracle) actively queries the teacher (i.e., the human oracle) about the label of the most informative examples (i.e., generated test cases).

\subsection{Maximizing the Probability to Label a Failing Test Case}\label{sec:learning}
In order to construct an effective automatic oracle, the \emph{class imbalance problem} (CIP) needs to be addressed. Usually, there are more test inputs that do \emph{not} expose the bug. In other words, a test generation technique normally generates more passing than failing test cases (whether or not we know the label of the test case). \tool addresses the CIP partially already using the mutational fuzzing approach to generate more test cases in the ``vicinity'' of the failing test case (\autoref{sec:fuzzing}). However, we still find that most generated inputs are passing. The human oracle $\mathcal{H}$ has limited time. So asking her to label mostly passing test cases is unproductive.

How do we maximize the probability that the human $\mathcal{H}$ is queried with mostly failing tests  (without knowing whether the test case  is \emph{actually} failing)? More importantly, how can we predict the probability that a test case is failing if the automatic oracle can only deliver hard decisions (passing \emph{or} failing)?

\begin{algorithm}[t]
\renewcommand{\algorithmicrequire}{\textbf{Input:}}
\renewcommand{\algorithmicensure}{\textbf{return}}
\caption{\textsc{decide2label} auxiliary function}\label{alg:label}
\begin{algorithmic}[1] 
\REQUIRE Unlabeled test case $t_\text{\textbf{?}}$, Automatic oracle $\mathcal{O}$
\REQUIRE Committee size $S$ 
\STATE Labeled test $t = \mathcal{O}(t_\text{\textbf{?}})$
\IF{$t$ is labeled as failing}
  \STATE \textbf{return} \texttt{true}
\ENDIF
\STATE $\text{\emph{votes}} = 0$
\FOR{Index $i$ from $1$ to $S$}
  \STATE Generated test $t'_\text{\textbf{?}} = \text{\textsc{fuzz}}(t_\text{\textbf{?}})$
  \STATE Labeled test $t'_\text{\cmark}\leftarrow$ assume $t'_\text{\textbf{?}}$ is labeled as passing
  \STATE Labeled test $t'_\text{\xmark}\leftarrow$ assume $t'_\text{\textbf{?}}$ is labeled as failing
  \STATE Hypothetical oracle $\mathcal{O}_\text{\cmark} = \text{\textsc{smt\_learn}}(T \cup \{t'_\text{\cmark}\})$
  \STATE Hypothetical oracle $\mathcal{O}_\text{\xmark} = \text{\textsc{smt\_learn}}(T \cup \{t'_\text{\xmark}\})$
  \STATE \textbf{if} $\mathcal{O}_\text{\cmark}$ labels $t_\text{\textbf{?}}$ as failing \textbf{then} increment $\emph{votes}$ \textbf{end if}
  \STATE \textbf{if} $\mathcal{O}_\text{\xmark}$ labels $t_\text{\textbf{?}}$ as failing \textbf{then} increment $\emph{votes}$ \textbf{end if}
\ENDFOR
\STATE Failure probability estimate $\hat\theta = \frac{\text{\emph{votes}}}{2\cdot S}$
\ENSURE $(\hat\theta \ge 0.5)$ \ \emph{// return true if the majority votes for} failing
\end{algorithmic}
\end{algorithm}

Algorithm~\ref{alg:label} provides an overview of how \textsc{decide2label} in Alg.~\autoref{alg:overview} maximizes the probability that the human oracle is asked to label mostly failing test cases. Firstly, if the current automatic oracle $\mathcal{O}$ predicts the label of the generated test $t_\text{\textbf{?}}$ as failing, then conservatively $t_\text{\textbf{?}}$ is passed on to the human oracle for labeling (i.e., \textsc{decide2label} returns \emph{true}; Lines~1--4). 

\textbf{Oracle committee}. Secondly, if the majority of an unbiased committee of oracles predicts the label of $t_\text{\textbf{?}}$ as failing, then it is passed on to the human oracle, as well. \tool constructs the oracle committee to answer the question how to predict the probability that $t_\text{\textbf{?}}$ is failing if our oracle $\mathcal{O}$ can only provide hard decisions. \tool effectively generates a given number $S$ of test cases by fuzzing $t_\text{\textbf{?}}$ and assigns each label $L\in \{\text{\cmark}, \text{\xmark}\}$ to each generated test case $t'_\text{\textbf{?}}$ (Lines~7--9). For each such hypothetically labeled test case $t'_L$, \tool generates an oracle $\mathcal{O}_L$ and asks that hypothetical oracle about the label of the original, unlabeled test case $t_\text{\textbf{?}}$. If the majority of hypothetical oracles predict the label of $t_\text{\textbf{?}}$ as failing, then \textsc{decide2label} returns \emph{true}, as well (Lines~12--16). Otherwise, \textsc{decide2label} returns \emph{false}.

The committee is \emph{unbiased} because \tool counts the vote from an oracle that is trained with the labeled test cases $T$ plus a random, new test case hypothetically labeled as failing with the same weight as the vote from an oracle that is trained with the \emph{same} new test case hypothetically labeled as passing. The key idea is to construct so-called \emph{look-one-ahead oracles} \cite{Holub2008} which are given \emph{all the actual evidence} (i.e., the labeled test cases $T$) \emph{plus one piece of hypothetical evidence}, (i.e., a random unlabled test case which is assigned a random label). Our experiments show, as more tests are labeled (i.e., $|T|$ increases), the prediction accuracy of the look-one-ahead oracles improves, too. To boost statistical power, \tool constructs and queries $2\cdot S$ such look-one-ahead oracles.

\section{Implementation}
\textsc{Learn2Fix} is implemented in Python v3.7 and builds on the \textsc{Incal} tool,\footnote{\url{https://github.com/ML-KULeuven/incal}} the \textsc{pywmi} toolbox for probabilistic inferencing (which includes a solver for SMT(LRA) constraints),\footnote{\url{https://pypi.org/project/pywmi/}} the \textsc{LattE} model counting tool,\footnote{\url{https://www.math.ucdavis.edu/~latte/software.php}} and the \textsc{NumPy} scientific computing library.\footnote{\url{https://numpy.org/}} Most of these libraries are dependencies that are inherited from \textsc{Incal}.

\begin{figure}[t]\centering
\includegraphics[width=0.5\textwidth]{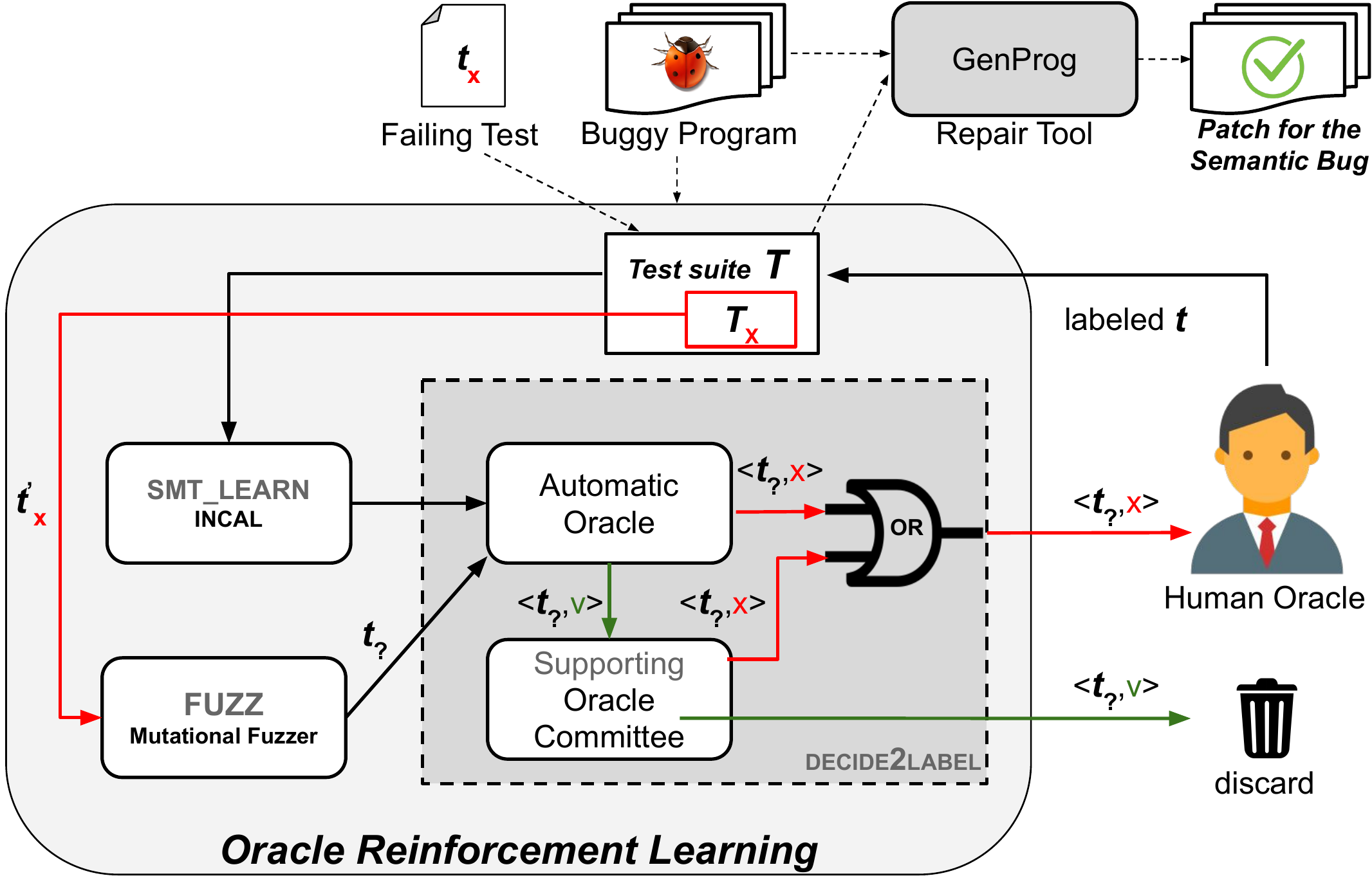}
\caption{Architecture and Implementation \textsc{Learn2Fix}}
\label{fig:diagram}
\end{figure}

Figure \ref{fig:diagram} gives an architectural perspective on \textsc{Learn2Fix}. At first the test suites $T$ and $T_\text{\xmark}$ are seeded with the \emph{failing test case} $t_\text{\xmark}$. The \textsc{smt\_learn} method uses \textsc{Incal} to learn the first \emph{automatic oracle} and the first \emph{oracle committee} from the labeled test suite. Meanwhile, the \textsc{fuzz} method implements the \emph{mutational fuzzer} which generates a new test case $t_\text{\textbf{?}}$.

Within the \textsc{decide2label} method, first the automatic oracle and then the supporting oracle committee are asked to predict the label of the generated test case. If either oracle or committee predict $t_\text{\textbf{?}}$ to be failing, $\langle t_\text{\textbf{?}},\text{\xmark}\rangle$,  $t_\text{\textbf{?}}$ is sent to the \emph{human oracle} for labeling. Otherwise, the generated test case is discarded. The test case that is labeled by the human oracle is added to the labeled test suite $T$. From the augmented test suite, \textsc{Incal} is used to construct an improved automatic oracle and oracle committee. The improved oracles are then used to make better predictions about which test cases to pass to the human oracle. This closes the \emph{active learning feedback loop}.

\textbf{\textsc{Incal}} \cite{incal} takes a \emph{variable domain} which specifies the type and number of the constrained variables, and a set of positive and negative examples. An \emph{example} is a vector of values for the variables in the domain that is specified. In our case, a positive example, which is supposed to satisfy the learned constraint, is a failing test case. From the domain and examples, \textsc{Incal} produces an SMT(LRA) constraint that is satisfied by all positive examples and by none of the negative examples. While our implementation of \tool is technically bound to SMT(LRA) constraints, we believe as the field of SMT learning advances more domains, such as strings, will become available to our approach.

The \textbf{mutational fuzzer} takes a test case and applies various mutation operators to generate a new test case. In \tool, a \emph{test case} $\langle \vec{i},o\rangle$ is a tuple consisting of a test input $\vec{i}$ and the actual output $o$. The mutational fuzzer extracts and mutates the test input to produce a new test input, and constructs the new test case by executing the new test input on the program (which produces the new actual output). As mutation operators, we use several simple operations, such as incrementing or decrementing the value, adding or subtracting ten, multiplying or dividing by ten, replacing the value with a random number. For each position in the test input, zero or more  operators can be applied.  Through the mutational fuzzer, it is possible to generate both passing and failing test cases. 

The \textbf{automatic oracle} is a first-order logic SMT formula generated by the SMT Learner. Given the program inputs and the actual output, the satisfiability of this formula predicts the label of the test case. Each time the human oracle labels a test case, the automatic oracle is being updated.

The \textbf{supporting oracle committee} contains a set of automatic oracles. Each automatic oracle is generated by the SMT learner taking the labeled test suite and one generated test case with a hypothetical label (Algorithm \ref{alg:label}; \autoref{sec:learning}). This generated test case is obtained from the mutational fuzzer. \tool constructs a new oracle committee every time a new test case is labeled and added to the labeled test suite $T$.

\section{Experimental Setup}
\subsection{Research Questions}\vspace{-0.1cm}
\begin{description}[itemsep=2pt,leftmargin=0.95cm]
  \item[\textbf{RQ.1 (Oracle quality)}.] How accurately does \tool's automatic oracle label manually constructed \& labeled (failing) test cases provided in the repair benchmark?
  \item[\textbf{RQ.2 (Human effort).}] What is the proportion of generated test cases that are sent to the human oracle for labeling? Does the probability to send mostly failing test cases indeed increase versus a random choice of test cases? 
  \item[\textbf{RQ.3 (Patch quality).}] How does the quality of patches produced using \tool's automatically generated test suite compare to the quality of patches produced using the manually constructed test suite that is provided with the repair benchmark? Specifically, how many subjects can be repaired and what is the proportion of validation (heldout) test cases that pass on the patched program?
\end{description}

\subsection{Experimental Subjects}\label{sec:subjects}\vspace{-0.1cm}
To evaluate \tool and answer the research questions, we chose the \emph{benchmark} according to the following criteria: 
\begin{enumerate}[itemsep=2pt]
  \item It should contain a sufficiently large number of programs that are algorithmically complex. 
  \item It should contain a diverse set of real defects that cause \emph{functional bugs}, i.e., programs produce unexpected output for some inputs. For each subject, there should be 1 bug.
  \item For each subject, there should be a \emph{golden version}, i.e., a slightly changed program where the bug has been fixed. To automate a large number of experiments, the golden version will \emph{act as the human oracle} $\mathcal{H}$. Any discrepancy between the output produced by the golden and buggy version for the same test input is labeled as failing test.
  \item For each subject, it should contain a \emph{manually constructed} and labeled \emph{training test suite} and a \emph{validation test suite}. Both test suites combined will be used to evaluate oracle quality while training and validation test suite will be used to generate a patch and evaluate patch quality, resp.
  \item For each subject, it should contain at least a \emph{failing test} in the training test suite, i.e., a test input for which the buggy and golden version produce different outputs. Otherwise, \tool cannot be started.
  \item For each subject, it should contain test inputs that have a constant number of numeric input values. For each such test input, the program should produce a numeric output. Otherwise, \textsc{Incal} cannot be used to learn the oracle $\mathcal{O}$.
\end{enumerate}
\textbf{Codeflaws} \cite{codeflaws} satisfies the first four selection criteria. Codeflaws consists of 3902 buggy programs and the corresponding golden versions. For each buggy program there exist manually constructed and labeled training and validation test suites. The programs originate from 1653 users of Codeforces where they competed in different programming contests and solved three to five programming problems. In their 2017 publication, the authors claim ``to our best knowledge, in automatic program repair evaluation, our benchmark has the largest number of real defects obtained from the largest number of subject programs to date'' \cite{codeflaws}.

From all subjects in Codeflaws, 552 subjects satisfy the last two criteria (cf.  \autoref{fig:subjects}). 2298 subjects took input files with more than one line, 323 took inputs that are \emph{not} numeric, 82 did \emph{not} contain functional bugs (but instead, e.g., crashes or timeouts), 39 did \emph{not} contain failing tests, one (1) subject had a non-constant number of input variables across test input files, and 607 subjects produced a non-numeric output.

IntroClass and ManyBugs benchmarks \cite{manybugs} do not satisfy our selection cri\-ter\-ia. ManyBugs consists of programs that take complex, non-numeric inputs which does not satisfy our sixth criterion. IntroClass consists of programs that implement one of six very simple functions (e.g., return the smallest of three numbers) which does not satisfy our first selection criterion.

\begin{figure}\footnotesize\centering
\begin{tabular}{@{}c@{ \ }c@{ }|@{ }l@{ \ }l@{}}
\textbf{\#} & \textbf{Defect} & \textbf{Description} & \textbf{Example}\\ 
\textbf{72} & DCCR & Replace constant & \texttt{- for(i=n+1;i<=90;i++)}\\
&&with variable/constant & \texttt{+ for(i=n+1;i<=100;i++)}\\\hline
\textbf{63} & OILN & Tighten condition& \texttt{- if(t\%2==0)}\\
&&or loosen condition & \texttt{+ if(t\%2==0 \&\& t!=2)}\\\hline
\textbf{59} & ORRN & Replace relational & \texttt{- if(sum>n)}\\
&& operator& \texttt{+ if(sum>=n)}\\\hline
\textbf{50} & HIMS & Insert multiple & \texttt{+ freopen("input.txt",}\\
&&non-branch statements& \quad\quad\quad\quad\quad\quad\texttt{"r", stdin);}\\
&&& \texttt{+ freopen("output.txt",}\\
&&& \quad\quad\quad\quad\quad\quad\texttt{"w", stdout);}\\\hline
\textbf{48} & OAIS & Insert/Delete  & \texttt{- max += days\%2;}\\
&&arithmetic operator& \texttt{+ max += (days\%7)\%2;}\\\hline
\textbf{48} & HOTH & Other higher order & \texttt{- scanf("\%s",h);}\\
&&defect classes& \texttt{+ for(i=0;i<71;i++)}\\
&&& \texttt{+ scanf("\%c",\&h[i]);}\\
\end{tabular}
\caption{\textbf{Top-5 defect classes} (out of 34) for our 552 Codeflaws subjects. First column (\#) lists the number of subjects belonging to that defect class.}
\label{fig:subjects}
\end{figure}

\textbf{GenProg}. As automatic repair tool, we chose GenProg \cite{genprog} because it is a mature tool that has been shown to repair large programs cost-effectively. GenProg is already set up with the Codeflaws repair benchmark.\footnote{We also sought to conduct experiments with the Angelix \cite{angelix} automatic repair tool. However, despite enlisting the help of the Angelix authors a few weeks before submission, we could not get Angelix to work with Codeflaws. It seems recent upstream changes in the dependencies introduced a compatibility issue: \url{https://github.com/mechtaev/angelix/issues/11} (DBR: Not our report).} 

\subsection{Setup and Infrastructure}\vspace{-0.1cm}
For each subject program, given a single failing input we run \textsc{Learn2fix} to produce a labeled test suite, including both failing and passing test cases. The detailed workflow for this step is shown in Figure~\ref{fig:diagram}. The data collected from the first step (i.e., test suite generation and labeling) is used to answer the first two research questions (RQ.1 and RQ.2). To answer RQ.3, we run GenProg twice with an attempt to repair the buggy subject using the labeled test suite generated by \textsc{Learn2fix} and the manually constructed training test suite, respectively.

\begin{figure*}\centering
\begin{minipage}{1\columnwidth}\centering
\includegraphics[width=\textwidth]{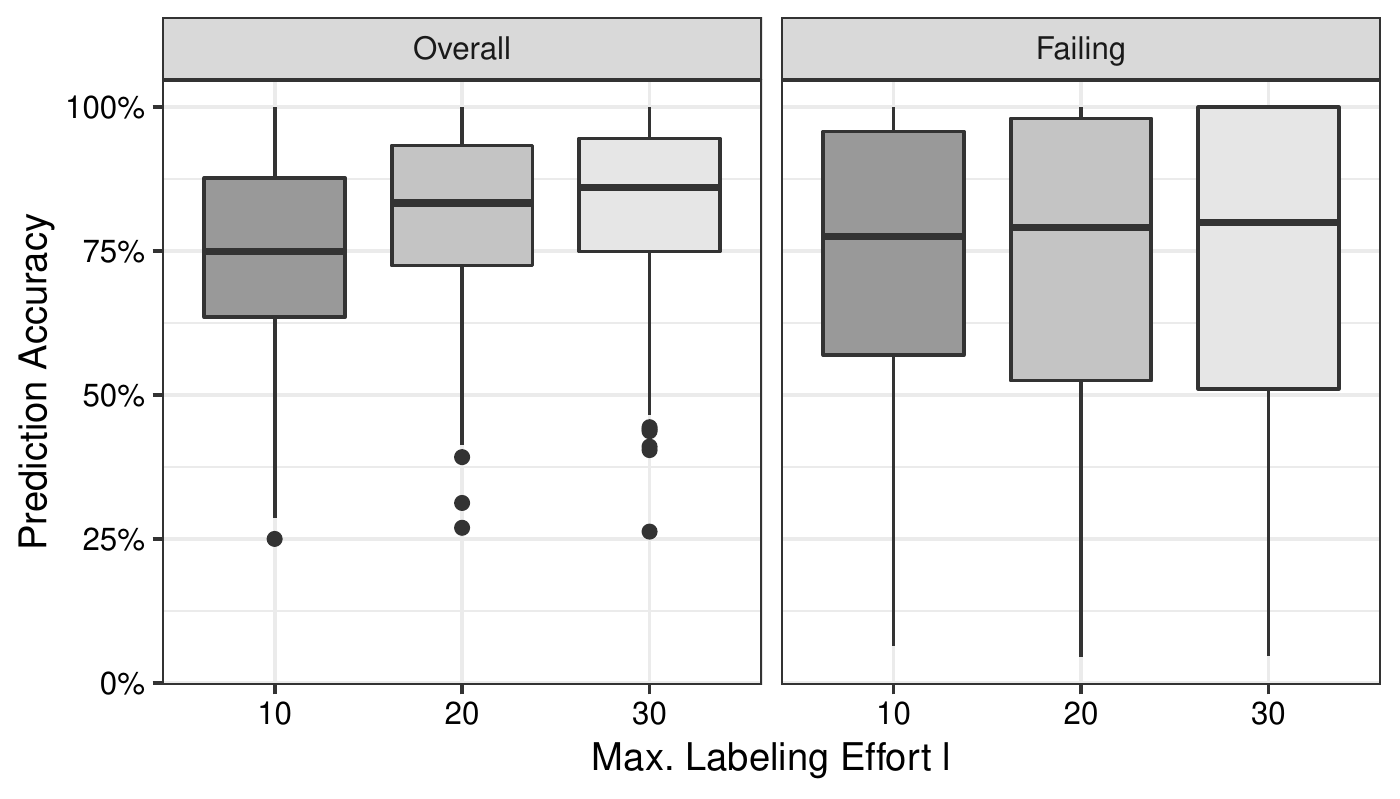}
\footnotesize (a)~Prediction Accuracy
\end{minipage} 
\begin{minipage}{0.8\columnwidth}\centering
\includegraphics[width=0.52\textwidth]{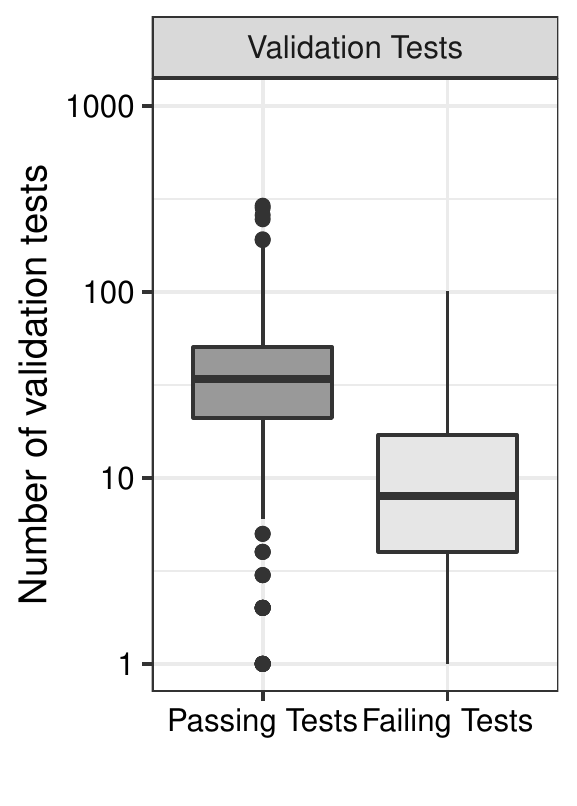}
\includegraphics[width=0.38\textwidth]{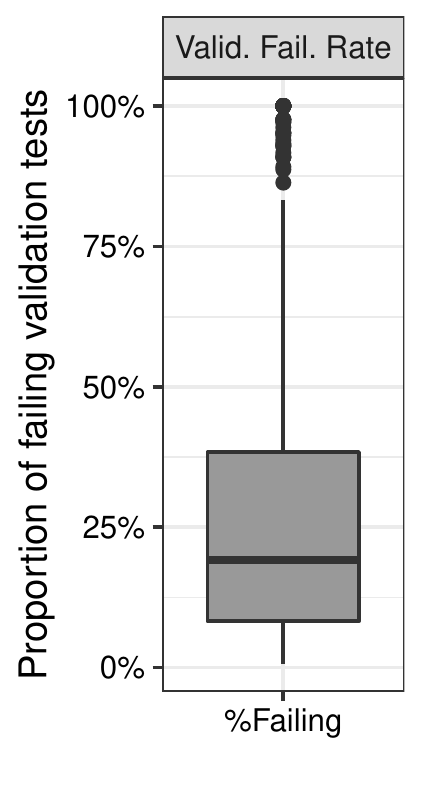} 
\footnotesize (b)~Validation Test Suites 
\end{minipage} 
\caption{\textbf{Oracle Quality}. \emph{On the left (a)}: Boxplots showing the accuracy of \tool's automatic oracle $\mathcal{O}$ when predicting the labels of manually constructed and labeled validation test cases in the repair benchmark, when varying the maximal number of labeling queries to the human oracle (i.e., $l$ in Algorithm~\ref{alg:overview}). For overall accuracy (a.left), the observed increases are statistically significant according to the \emph{paired one-sided Wilcoxon test} ($p<0.0001$). For conditional accuracy (a.right), there is no statistically significant difference ($p>0.05$).  
\emph{On the right (b)}: Boxplots showing some statistics about the validation test suites, including the distribution of passing and failing tests  across subjects. \vspace{-0.2cm}}
\label{fig:oracleQuality} 
\end{figure*}

\newpage
\noindent
For our experiments, we fixed the following values.

\begin{itemize}
  \item \emph{Timeouts.} We set the maximal time for auto-generating a labeled test suite (Alg. \ref{alg:overview}) as well as for auto-generating a patch to 10 minutes each.
  \item \emph{Committee size.} We set the size of the oracle committee to 20 members (i.e., $S=10$ in  Alg.~\ref{alg:label}).
  \item \emph{Maximal labeling effort}. We set the maximal number of queries to the human oracle to 10, 20, and 30, repeating the entire set of experiments for all three parameters (i.e., $l\in \{10,20,30\}$ in Alg.~\ref{alg:overview}).
\end{itemize} 

\emph{Experiment Repetition}. To mitigate the impact of randomness and to gain statistical power for the experimental results, we repeat each experiment 30 times.

\emph{Infrastructure}. We evaluated our approach within a Docker container that has access 64GB of main memory and 32 logical Intel Haswell processor cores (each at 2.0GhZ) and that runs the Ubuntu 16.04.03 LTS (Xenial Xerus) 64-bit OS.\vspace{-0.1cm}

\subsection{Reproducibility}\vspace{-0.05cm}
To facilitate open science and reproducibility, we make our implementation of \tool, our collected data, and scripts available at first in a blinded manner for the reviewer's scrutiny and, upon acceptance, publically for everyone to reproduce our results. We provide concrete  instructions for reproducing the experiments and all artifacts here:
\begin{itemize}
\item[$\triangleright$]\textcolor{blue}{\url{https:/github.com/mboehme/learn2fix}}
\end{itemize}

\section{Experimental Results}
\subsection*{RQ.1 Test Oracle Quality}\vspace{-0.05cm}
\emph{Validation tests}. The Codeflaws benchmark \cite{codeflaws} comes with a large number of manually labeled test cases (i.e., training and heldout test cases). We use these validation tests to check the accuracy of \tool's automatic oracles. In \autoref{fig:oracleQuality}.b, we see the distribution of the number of passing and failing validation tests across all subjects. On average, each subject is accompanied by 42 passing and 13 failing validation tests.

\emph{Measures of oracle quality}. The classical \emph{prediction accuracy} is computed as the proportion of validation tests for which both human and automatic oracle agree. However, due to the class imbalance problem, classical accuracy is not a good measure of oracle quality. If about 90\% of the generated tests are actually passing, even a low-quality oracle that predicts \emph{all} tests as passing would have a prediction accuracy of 90\%. To address the class imbalance problem, we also report the \emph{conditional accuracy} which is computed as the proportion of \emph{actually failing} validation tests that the automatic oracle also labels as failing (i.e., conditioned on the minority class).

\emph{Result presentation}. Throughout the results section, the box plot is our main means of presenting the data. For each subject, labeling effort, and measure, we computed average values over all 30 runs. For instance, when configured with a maximal labeling effort of $l=10$ queries, \tool's prediction accuracy for the subject \texttt{1-A-bug-18353198-18353306} is 83.4\% when averaged over all 30 runs. A \emph{boxplot} shows several interesting statistics for the distribution of this average value \emph{over all subjects}. For instance, from the first boxplot in \autoref{fig:oracleQuality}.a, we can derive that for the middle 50\% of the 552 subjects, \tool exhibits a prediction accuracy between 63\% and 88\% with a median of 75\%. For a quarter of subjects the prediction accuracy is higher than 88\% (up 100\%).

\result{Even though \tool has only ever seen a single failing test case from the manually labeled validation test suite, the automatic oracle is able to accurately predict the label of 75--84\% validation tests for the median subject (Fig.~\ref{fig:oracleQuality}.a).

\ \ The prediction accuracy further increases as more queries can be sent to the human oracle. For instance, if the human is willing to label 30 tests, the median prediction accuracy is 15\% higher than if the human only wants to label ten.

\ \ Even if we focus only on the minority class, Fig.~\ref{fig:oracleQuality} shows that \tool's automatic oracle labels \emph{at least} 78\% of failing validation tests correctly for the majority of subjects (i.e., the median conditional accuracy is 78\% or higher).
}
\textbf{Result}. \emph{\tool produces high-quality test oracles that improve with the number of queries to the human oracle.}

\begin{figure*}\centering
\begin{minipage}{\columnwidth}\centering
\includegraphics[width=0.85\textwidth]{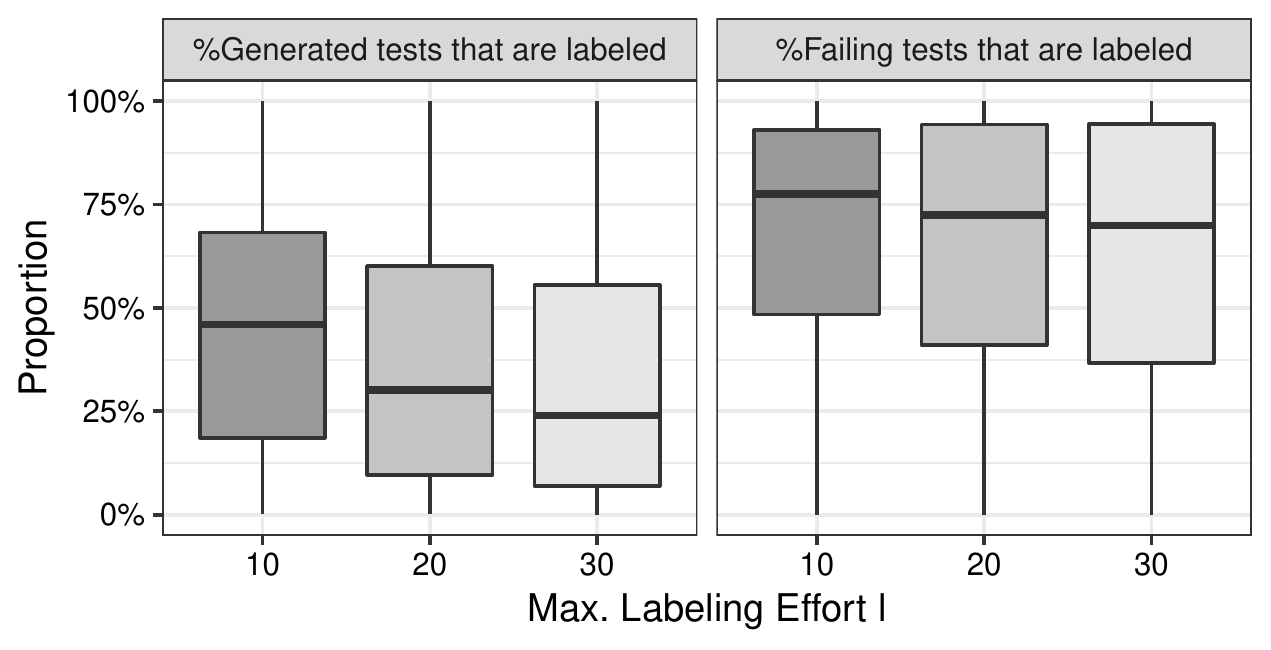}
\footnotesize (a)~Proportion of generated / failing tests that are labeled across all subjects. 
\end{minipage}\quad
\begin{minipage}{\columnwidth}\centering
\includegraphics[width=0.85\textwidth]{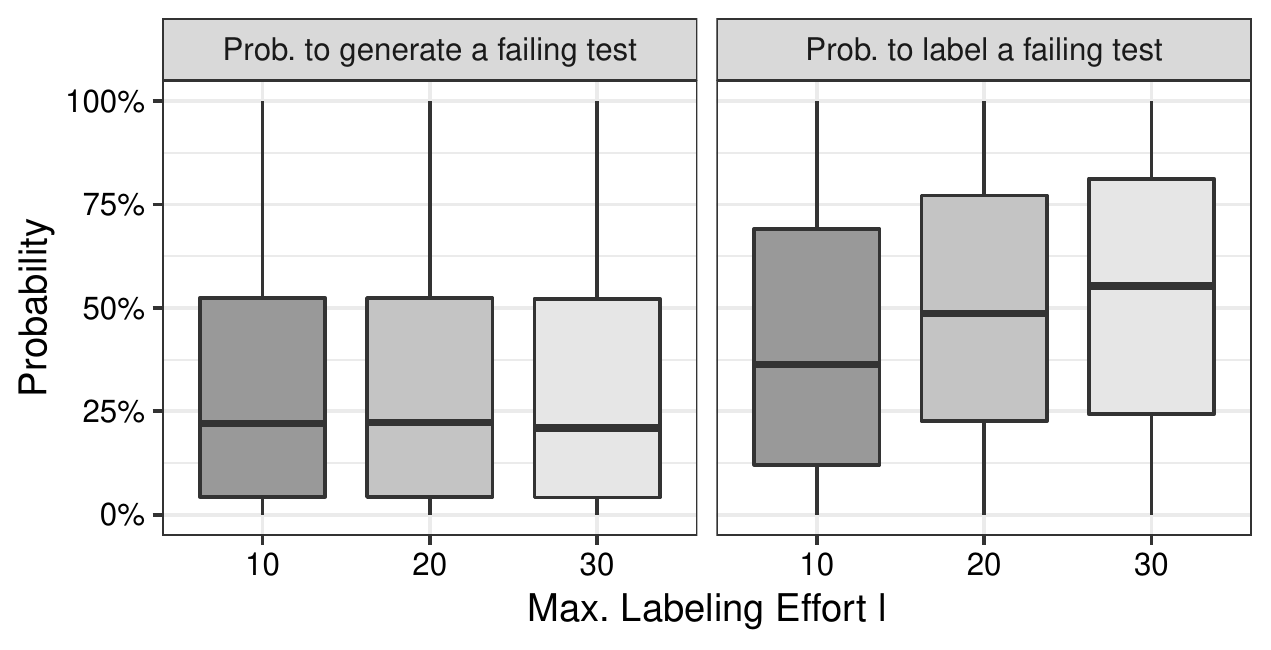}
\footnotesize (b)~Probability to generate / label a failing test case across all subjects.
\end{minipage} 
\caption{\textbf{Human Effort}.  \emph{On the left (a)}: Boxplots showing the proportion of generated or failing tests that \tool sends to the human oracle for labeling. All observed decreases are statistically significant (\emph{paired one-sided Wilcoxon test}; $p<0.01$).
\emph{On the right (b)}: Boxplots showing the probability~that~\tool sends a failing test relative to the probability that \tool generates a failing test. Observed increases (b.right) are statistically significant ($p<0.0001$).\vspace{-0.4cm}}
\label{fig:humanEffort} 
\end{figure*}
 
\subsection*{RQ.2 Human Effort}\vspace{-0.1cm}
\emph{Measures of human effort}. We report two metrics. Firstly, we measure the \emph{proportion of generated tests that are labeled} given the human is willing to label at most $l\in \{10,20,30\}$ generated tests. That is, we abort the test generation once the human has labeled $l$ tests. The lower the proportion of labeled generated test cases, the lower the human effort.

Secondly, we measure the \emph{probability to label a failing test}, i.e., how likely it is that the human oracle labels a received test as failing. The key objective of \tool is to maximize the probability that the human oracle is labeling failing test cases, i.e., the minority class. If the probability to \emph{label} a failing test case is much greater than the probability to \emph{generate} a failing test case, the human effort is reduced.

\result{\vspace{-0.1cm}\emph{As the (conditional) prediction accuracy of the automatic oracle increases, the proportion of generated tests that are labeled \emph{decreases} (Fig.~\ref{fig:humanEffort}.a-left). Meanwhile, the proportion of generated test cases that are labeled \emph{as failing} remains above 70\% (Fig.~\ref{fig:humanEffort}.a-right). In other words, while a smaller proportion of generated test cases require human scrutiny, \tool retains roughly the same proficiency to identify failing test cases to send to the human oracle.}}\vspace{-0.1cm}
For the median subject, almost every second test case is sent for labeling if the human oracle is willing to label only ten test cases ($l=10$). However, for the median subject, only one in four generated tests is sent for labeling if the human is willing to label three times more test cases ($l=30$). Recall that no more tests are generated once the human oracle has labeled $l$ generated test cases.

\result{\vspace{-0.1cm}\emph{As the (conditional) prediction accuracy of the automatic oracle increases, the probability that a test--that is labeled by the human oracle--turns out failing \emph{increases} (Fig.~\ref{fig:humanEffort}.b-right). Meanwhile, the probability to generate a failing test (unsurprisingly) remains about the same (Fig.~\ref{fig:humanEffort}.b-left). In fact, for the median and average subject, when $l=30$, the probability to label a failing test case is twice and 11 times higher than expected by random labeling, respectively.}}\vspace{-0.1cm}
For the median subject, about a quarter of the generated tests are actually failing. However, if the human is willing to label $l=30$ generated tests, for the median subject, the probability that \tool labels a failing test is about twice as high. 

\textbf{Result}. \emph{\tool reduces human effort versus random labeling. As oracle quality improves, effort is further reduced.}

\begin{figure*}\centering
\begin{minipage}{0.85\columnwidth}\centering
\includegraphics[width=\textwidth]{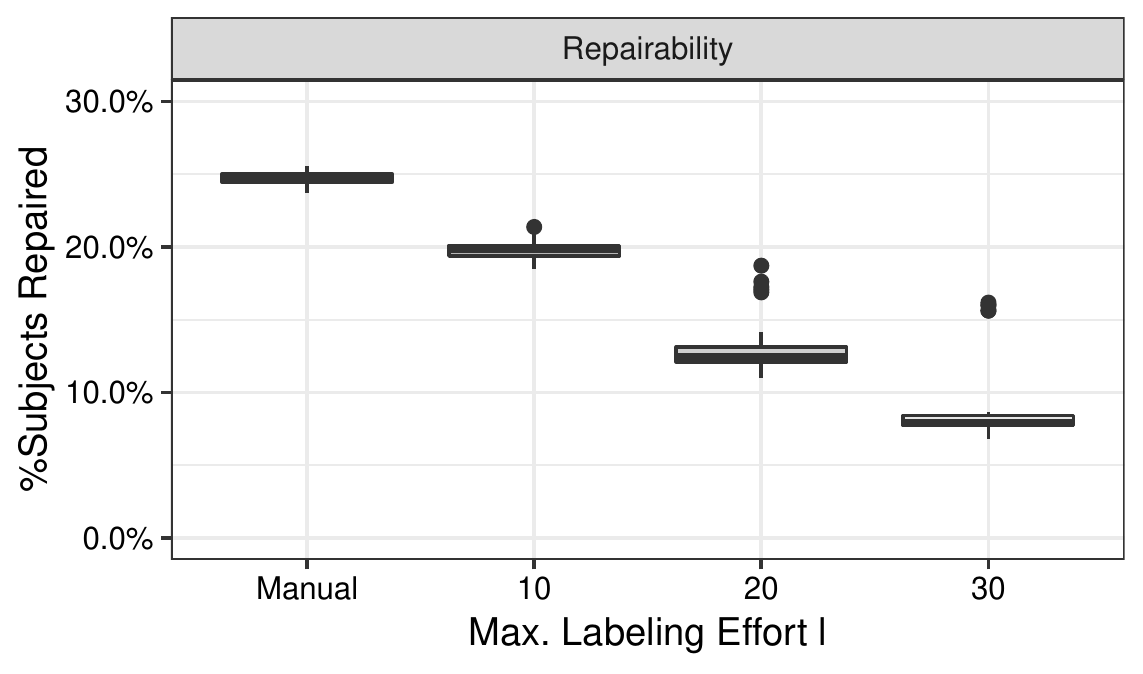}
\end{minipage} \ 
\begin{minipage}{0.29\columnwidth}\centering
\includegraphics[width=\textwidth]{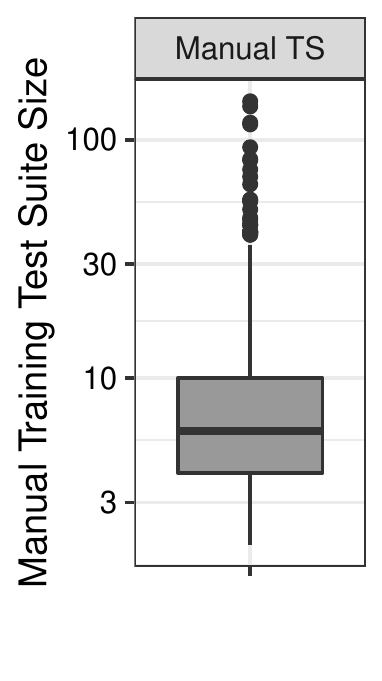}
\end{minipage} \ 
\begin{minipage}{0.85\columnwidth}\centering
\includegraphics[width=\textwidth]{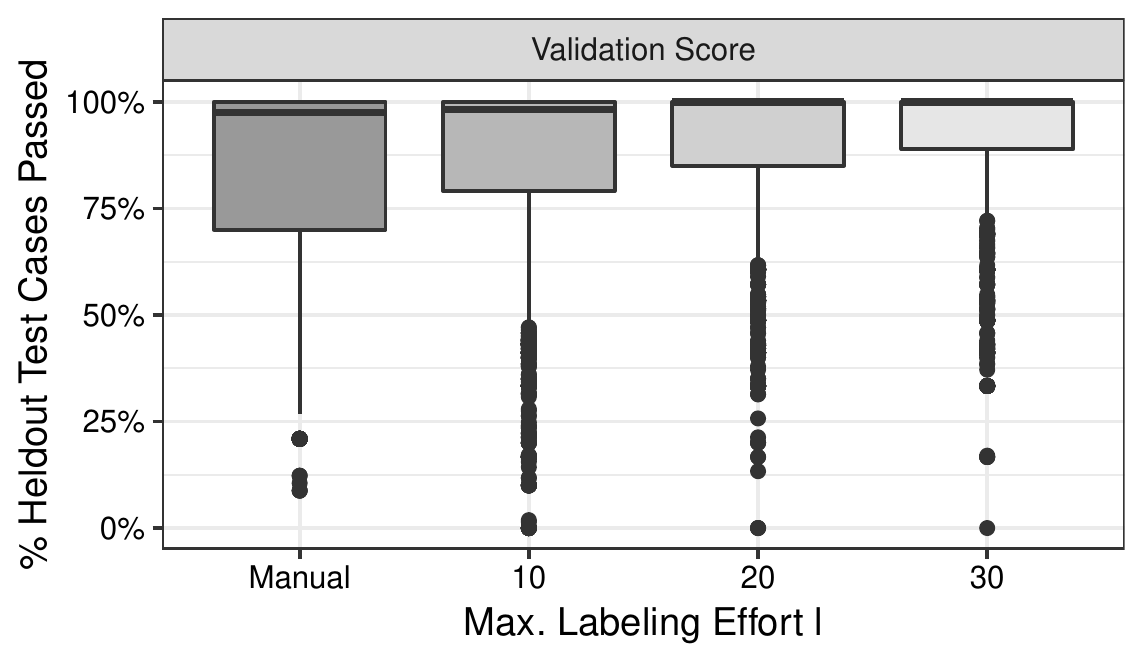}
\end{minipage} 
\caption{\textbf{Patch Quality}.
\emph{On the left}: The proportion of subjects that could be successfully repaired  (across all 30 runs), i.e., a patch was generated such that all (generated) \emph{training} tests pass. 
\emph{In the middle}: The number of manually constructed \emph{training} test cases as provided with the Codeflaws repair benchmark (across all 552 subjects).
\emph{On the right}: The proportion of manually constructed \emph{validation} test cases that pass on the patched program (across all 552 subjects). According to the \emph{unpaired, two-sided Wilcoxon test}, the observed increases between Manual and $l=20$ are statistically significant ($p<0.001$). There is no statistically significant difference from $l=20$ to $l=30$ ($p>0.05$) In both cases, a mean of 90\% heldout tests pass on the patches produced by \tool.
}
\label{fig:patchQuality} 
\end{figure*}
\subsection*{RQ.3 Patch Quality}
\emph{Training and validation}. For each subject, the Codeflaws repair benchmark provides manually constructed training and validation test suites. Given the \emph{training test suite}, GenProg attempts to produce a patch such that all test cases in the training set are passing. In addition, for each subject \tool auto-generates three training test suites of size $l\in \{10,20,30\}$. We assume that the human oracle provides the expected output for the labeled tests. We call the manually constructed training suite as \texttt{manual} and \tool's training suites as \texttt{10}, \texttt{20}, or \texttt{30} depending on suite's size $l$. The \emph{validation test suite} is used to measure the quality of all four patches.

\emph{Measures of patch quality}. Firstly, we measure \emph{repairability} as the proportion of subjects that were successfully repaired in at least one of three runs. In other words, for those subjects all tests in the provided \emph{training} test suite pass on the patched program. Secondly, we measure the \emph{validation score} as the proportion of validation tests that pass on the patched program. Primarily, we would like to maximize the validation score as it also measures overfitting (unlike repairability).

\result{\vspace{-0.1cm}\emph{While fewer subjects are repaired when GenProg is given \tool's training suite, the validation score is higher. As the size of the training suite increases, the repairability decreases: Fewer repairs are produced (Fig.~\ref{fig:patchQuality}.left). Hence, we explain the increased repairability--when given the manual suite--by the lower number of training tests (Fig.~\ref{fig:patchQuality}.middle). The median and average number of manual training tests per subject is 6 and 10.8 test cases, respectively. %
%
The validation score of the patches that are produced with \tool's training tests \emph{increases} as test suite size increases (Fig. \ref{fig:patchQuality}.right): Our auto-generated patches are better. The proportion of validation test cases that still fail on the generated repair is 31\% smaller, on the average, when the repair is produced by \tool ($l=30$) than by the manually constructed training suite (cf. \autoref{fig:patchquality2}).}\vspace{-0.1cm}}\vspace{-0.1cm}
 
\textbf{Result}. \emph{\tool produces test suites that are well-suited for automatic program repair to produce high-quality patches.}

\begin{figure}[t]\small\centering
\begin{tabular}{l|rr}
&\textbf{Median} & \textbf{Average}\\\hline
Manually generated test suite & 2.4\% & 14.9\% \\ 
Auto-generated test suite ($l=10$) & 1.6\% & 13.8\% \\
Auto-generated test suite ($l=20$) & 0\% & 10.1\% \\
Auto-generated test suite ($l=30$) & 0\% & 10.3\% \\ \hline
\end{tabular}\vspace{0.2cm} 
\caption{Proportion of validation tests that are still failing on the repaired subject (across all subjects). Complement of the validition score ($1-\text{\emph{score}}$).}
\label{fig:patchquality2}
\end{figure}
 
\section{Threats to Validity}
As for any empirical study, there are various threats to the validity of our results and conclusions. 
The first concern is \emph{external validity}, i.e., the degree to which our findings can be generalized to and across other subjects and tools. First, our results may not hold for other subjects. Particularly, our SMT learning tool \textsc{Incal} \cite{incal} works only for numeric variables. Hence, our subject programs were required to take numeric input values and produce numeric output values. However, we chose our subjects such that we had a large number of real, arithmetically complex, buggy programs containing diverse types of defects (i.e., 552 programs containing 34 defect types; cf. Sec. \ref{sec:subjects}).
Second, the results on patch quality (RQ.3) may turn out differently for automatic program repair (APR) tools other than GenProg. However, the APR tool used in our experiments, GenProg \cite{genprog}, is state-of-the-art and has been shown to repair large open-source programs cost-effectively.

The second concern is \emph{internal validity}, i.e., the degree to which our study minimizes  systematic error. Firstly, to mitigate spurious observations due to the randomness of the mutational fuzzer or the SMT learner and to gain statistical power, we repeated each experiment 30 times and report average values. Second, like implementations of other techniques, our tool may not faithfully implement \tool as presented in Algorithms~\ref{alg:overview} and \ref{alg:label}. However, to facilitate scrutiny and reproducibility, we make source code and all data available.

The third concern is \emph{construct validity}, i.e., the degree to which a test measures what it claims, or purports, to be measuring. To mitigate this threat, we motivate and discuss at least two measures for each of the three independent variables: oracle quality, human effort, and patch quality.

\section{Related Work}
This paper was inspired by the work of Holub et al. \cite{Holub2008} on an active learning approach to reduce human (labeling) effort during image classification. Faced with large amounts of unlabeled data, the authors required an efficient human labeling approach. Holub et al. proposed to send in each iteration to the human oracle only the \emph{most informative unlabeled point} (MUIP), i.e., that image where the classifier is most uncertain about its label. ``At first glance this may seem counter-intuitive: how can the algorithm know whether a group of unlabeled images will be informative, when, by definition, there is no label directly associated with the images'' \cite{Holub2008}? 

We extended the approach of Holub et al. by addressing the \emph{class imbalance problem} (generated tests are likely passing) which violates their assumption of equal probabilities over all labels. \tool conservatively sends all test cases to the human oracle that are predicted as likely failing. Moreover, we extended their pool-based approach, where the data have a fixed number of points, into a \emph{stream-based approach}, where data points are continuously generated and decided upon. Our insight is that a reliable probability estimate for a point's label can be derived with a fixed-size random classifier committee.

``The automation of test oracles is probably one of the most difficult problems\,in\,software\,testing'' according to Briand~\cite{briand}. There are two kinds of automatic test oracles. Barr et al. \cite{Barr2015a} provide a recent survey on the oracle problem in general while Pezz\`e and Chang \cite{automatedoracles} review the current state-of-the-research for constructing and evaluating automated test oracles specifically.
\emph{Implicit oracles} are compiler-induced and observable as crashes or timeouts \cite{Barr2015a}. Functional or logical bugs cannot be detected by implicit oracles. Implicit oracles can also be \emph{injected} \cite{asan,msan}. For instance, ASAN \cite{asan} induces a crash for inputs that expose a memory safety error. 
\emph{Explicit oracles} detect functional and other bugs and must be manually added. Typically, developers write assertions as explicit oracles \cite{assertions}. However, while assertions are added proactively, our goal is to construct an explicit oracle retroactively, i.e., the oracle should identify new test cases exposing a known error.

Bowring et al. \cite{bowring} also propose to train an oracle by queries to the human. Suppose, a test input $t$ was labeled as failing. The test oracle would assign an unlabeled test input which executes the same code with similar frequency as $t$ the same label as $t$. Unlabeled tests that execute other code or the same code with a different frequency as previously labeled tests are sent to the human for labeling. This induces a query for each increase in coverage.
In their study, even at 350 queries, the authors did \emph{not} observe a difference in prediction accuracy compared to a passive learning approach, where the classifier is trained with a random set of the same number of labeled tests.
In contrast, \tool requires no source code and addresses our key challenge, i.e., to minimize the number of queries while maximizing oracle quality, which we empirically demonstrate even at 10 queries.  Also, \tool realizes our larger vision of human-in-the-loop automatic program repair.

Other existing approaches \emph{infer a specification} of a program as regression oracle \cite{automatedoracles}---while the objective of \tool is to infer a bug oracle. A \emph{regression oracle} checks whether a new program version behaves like the previous version for unseen inputs. Machine learning can be used to learn the function between the program inputs and outputs. The learned function can be implicitly represented as classifier \cite{Hu2008,Braga2018,annoracle} or explicitly as a program assertions or likely invariants \cite{daikon}. Now, given a new program version, if for an input the output that is predicted by the regression oracle differs from the actual output, a bug may have been introduced in the recent changes.

While such existing techniques learn to identify the \emph{expected} behavior of a program, our \emph{bug oracle} learns to identify the \emph{unexpected} behavior (i.e., a known functional bug). We believe that this smaller task is achieved much more efficiently. For instance, in our motivating example (Sec. \ref{sec:motivating}), \tool does \emph{not} learn how to classify a triangle. Instead, it learns which triangles Steve's program classifies incorrectly. Groce et al. \cite{groce} present an approach to learn an automatic bug oracle for a classifier that has been observed to misclassify. However,  their approach uses properties that are specific to classifier programs. In contrast, \tool generally applies to all programs (as long as their domain and range can be represented in an existing satisfiability modulo theory).

To reduce the time it takes the human oracle to label tests in a generated test suite as passing or failing, several approaches have been proposed. Harman et al. \cite{oraclenumber} propose to \emph{reduce the number of tests generated} without compromising code coverage achieved. Afshan et al. \cite{readable} proposed to \emph{improve the human readability} of the generated test cases by incorporating a natural language model into the test generation process. McMinn et al. \cite{McMinn2010} propose more generally to \emph{extract oracle knowledge from various sources}, such as programmers, source code and documentation. Pastore et al. \cite{crowdoracles} propose to \emph{crowd source the labeling task} to a crowd of users on the Internet to label the test cases. Staats et al. \cite{Staats2012} suggest to \emph{identify a smaller subset of variables} to be checked that have the highest impact on the program output. However, none of these works studies the problem of constructing an automatic oracle that learns to label a continuous stream of generated test cases. 

The use of constraints as oracles has been explored within formula-based debugging \cite{fbdebug}. Given an input that fails on the current version and passes in a previous version, DARWIN \cite{darwin} constructs an SMT constraint that is satisfied by all inputs that pass on the current version and differ marginally from the failing input in their control flow behavior. This difference in control flow points to the location of the fault. Ermis et al. \cite{errorinvariants} intro\-duce the \emph{error invariant} which is anchored at a particular program location and represents variable values that still lead to an observed assertion violation. Angelic Debugging \cite{angelic} infers SMT constraints as statement-level specification for the faulty statement. None of the existing techniques produce an automatic oracle that can decide whether a test case does or does not expose a known functional error.

The generation of further failing test cases from one failing test has previously been explored. The AFL fuzzer \cite{aflfast,aflgo,aflsmart} supports a crash-mode for this purpose. More tests can also be generated to isolate the fault \cite{bugex}, or to improve auto-generated patches \cite{fix2fit}. Given only a stack trace, there exist techniques that generate a crash-reproducing input \cite{f3,hercules}. Unlike in our work, in all cases an automatic oracle is already assumed.
\section{Discussion and Future Work}
We envision a semi-automatic approach that negotiates the condition under which the bug is observed before repairing the bug. Strategically, the user is asked: ``\emph{For this other input, the program produces that output; is the bug observed}''? While the user might not have the expertise to understand the source code or to produce a patch, it seems reasonable to ask to distinguish expected from unexpected program behavior. Iteratively, an \emph{automatic bug oracle} is trained to predict the user's responses with increasing accuracy. Using the trained oracle, the user can be asked even more strategically. The key challenge that we addressed in this paper was to maximize the oracle's accuracy, given only a limited number of queries.

We presented \tool which realizes our approach for programs that take numeric inputs. We empirically showed that the quality of the trained oracles and auto-generated patches is reasonably high despite a relatively low labeling effort. The restriction to numeric inputs is inherited from \textsc{Incal} \cite{incal} a very recent (2018) technique that learns an SMT(LRA) constraint from positive and negative examples. Yet, programs that take arrays, strings, or structured objects require more sophisticated theories (e.g., \cite{hampi,boolector}). Some limitations are shared with symbolic execution which builds on SMT, as well. Realizing our approach for non-numeric inputs is left for future work.

We note that the learned constraint \emph{cannot} serve as ground truth, e.g., during semantic repair \cite{angelix,semfix}. The inferred oracle is only approximate and may itself mispredict. In future, we plan to explore the \emph{true risk} \cite{shalev} (i.e., the true probability of mis\-pre\-dic\-tion) within a probabilistic or statistical framework \cite{mcpa,stads}. Exact learning might be possible with equivalence queries \cite{angluin} in addition to the membership queries to the (human) teacher. As a more general approach, we also plan to evaluate binary classification with abstention \cite{abstention}. Only test inputs with ``uncertain'' label are passed to the human.

We are excited by the prospect of the first fully automatic end-to-end software debugging technique that starts with a user reporting a bug and ends with a patch that fixes the bug, where the patch has a better quality than one which is auto-generated from manually constructed test cases. Of course, there are abundant opportunities for future work, but our current results are already very encouraging.

\section*{Acknowledgements}
We thank Aldeida Aleti for the interesting discussions about this paper and her inspiring and truely insightful perspective on human bias in machine learning. We thank the anonymous reviewers for their constructive feedback. This research was fully or partially funded by the Australian Government through an Australian Research Council (ARC) Discovery Early Career Researcher Award (DE190100046).


\begin{thebibliography}{10}
\providecommand{\url}[1]{#1}
\csname url@samestyle\endcsname
\providecommand{\newblock}{\relax}
\providecommand{\bibinfo}[2]{#2}
\providecommand{\BIBentrySTDinterwordspacing}{\spaceskip=0pt\relax}
\providecommand{\BIBentryALTinterwordstretchfactor}{4}
\providecommand{\BIBentryALTinterwordspacing}{\spaceskip=\fontdimen2\font plus
\BIBentryALTinterwordstretchfactor\fontdimen3\font minus
  \fontdimen4\font\relax}
\providecommand{\BIBforeignlanguage}[2]{{%
\expandafter\ifx\csname l@#1\endcsname\relax
\typeout{** WARNING: IEEEtran.bst: No hyphenation pattern has been}%
\typeout{** loaded for the language `#1'. Using the pattern for}%
\typeout{** the default language instead.}%
\else
\language=\csname l@#1\endcsname
\fi
#2}}
\providecommand{\BIBdecl}{\relax}
\BIBdecl

\bibitem{survey1}
M.~Monperrus, ``Automatic software repair: A bibliography,'' \emph{ACM Comput.
  Surv.}, vol.~51, no.~1, pp. 17:1--17:24, Jan. 2018.

\bibitem{survey2}
L.~Gazzola, D.~Micucci, and L.~Mariani, ``{Automatic Software Repair: A
  Survey},'' \emph{IEEE Transactions on Software Engineering}, vol.~45, no.~1,
  pp. 34--67, 2019.

\bibitem{genprog}
C.~{Le Goues}, T.~V. Nguyen, S.~Forrest, and W.~Weimer, ``{GenProg: A generic
  method for automatic software repair},'' \emph{IEEE Transactions on Software
  Engineering}, vol.~38, no.~1, pp. 54--72, 2012.

\bibitem{incal}
S.~Kolb, S.~Teso, A.~Passerini, and L.~{De Raedt}, ``{Learning SMT(LRA)
  constraints using SMT solvers},'' \emph{IJCAI International Joint Conference
  on Artificial Intelligence}, vol. 2018-July, pp. 2333--2340, 2018.

\bibitem{dart}
P.~Godefroid, N.~Klarlund, and K.~Sen, ``Dart: Directed automated random
  testing,'' in \emph{Proceedings of the 2005 ACM SIGPLAN Conference on
  Programming Language Design and Implementation}, ser. PLDI '05, 2005, pp.
  213--223.

\bibitem{klee}
C.~Cadar, D.~Dunbar, and D.~Engler, ``Klee: Unassisted and automatic generation
  of high-coverage tests for complex systems programs,'' in \emph{Proceedings
  of the 8th USENIX Conference on Operating Systems Design and Implementation},
  ser. OSDI'08, 2008, pp. 209--224.

\bibitem{jpf}
C.~S. P\u{a}s\u{a}reanu and N.~Rungta, ``Symbolic pathfinder: Symbolic
  execution of java bytecode,'' in \emph{Proceedings of the IEEE/ACM
  International Conference on Automated Software Engineering}, ser. ASE '10,
  2010, pp. 179--180.

\bibitem{semfix}
H.~D.~T. Nguyen, D.~Qi, A.~Roychoudhury, and S.~Chandra, ``Semfix: Program
  repair via semantic analysis,'' in \emph{Proceedings of the 2013
  International Conference on Software Engineering}, ser. ICSE '13, 2013, pp.
  772--781.

\bibitem{angelix}
S.~Mechtaev, J.~Yi, and A.~Roychoudhury, ``{Angelix: Scalable multiline program
  patch synthesis via symbolic analysis},'' \emph{Proceedings - International
  Conference on Software Engineering}, pp. 691--701, 2016.

\bibitem{Holub2008}
A.~Holub, P.~Perona, and M.~C. Burl, ``{Entropy-based active learning for
  object recognition},'' \emph{2008 IEEE Computer Society Conference on
  Computer Vision and Pattern Recognition Workshops, CVPR Workshops}, pp. 1--8,
  2008.

\bibitem{codeflaws}
{Shin Hwei Tan}, {Jooyong Yi}, {Yulis}, S.~{Mechtaev}, and A.~{Roychoudhury},
  ``Codeflaws: a programming competition benchmark for evaluating automated
  program repair tools,'' in \emph{2017 IEEE/ACM 39th International Conference
  on Software Engineering Companion (ICSE-C)}, 2017.

\bibitem{plausible}
Z.~Qi, F.~Long, S.~Achour, and M.~Rinard, ``An analysis of patch plausibility
  and correctness for generate-and-validate patch generation systems,'' in
  \emph{Proceedings of the 2015 International Symposium on Software Testing and
  Analysis}, ser. ISSTA 2015, 2015, pp. 24--36.

\bibitem{overfitting}
E.~K. Smith, E.~T. Barr, C.~Le~Goues, and Y.~Brun, ``Is the cure worse than the
  disease? overfitting in automated program repair,'' in \emph{Proceedings of
  the 2015 10th Joint Meeting on Foundations of Software Engineering}, ser.
  ESEC/FSE 2015, 2015, pp. 532--543.

\bibitem{russcon}
R.~Williams, ``Triangle classification problem,''
  \url{https://russcon.org/triangle_classification.html}, July 2002.

\bibitem{fix2fit}
X.~Gao, S.~Mechtaev, and A.~Roychoudhury, ``Crash-avoiding program repair,'' in
  \emph{Proceedings of the 28th ACM SIGSOFT International Symposium on Software
  Testing and Analysis}, ser. ISSTA 2019, 2019, pp. 8--18.

\bibitem{synthesis}
S.~Jha, S.~Gulwani, S.~A. Seshia, and A.~Tiwari, ``Oracle-guided
  component-based program synthesis,'' in \emph{Proceedings of the 32Nd
  ACM/IEEE International Conference on Software Engineering - Volume 1}, ser.
  ICSE '10, 2010, pp. 215--224.

\bibitem{classimbalance}
N.~Japkowicz and S.~Stephen, ``The class imbalance problem: A systematic
  study,'' \emph{Intelligent data analysis}, vol.~6, no.~5, pp. 429--449, 2002.

\bibitem{manybugs}
C.~{Le Goues}, N.~Holtschulte, E.~K. Smith, Y.~Brun, P.~Devanbu, S.~Forrest,
  and W.~Weimer, ``The {ManyBugs} and {IntroClass} benchmarks for automated
  repair of {C} programs,'' \emph{IEEE Transactions on Software Engineering
  (TSE)}, vol.~41, no.~12, pp. 1236--1256, December 2015.

\bibitem{briand}
L.~C. {Briand}, ``Novel applications of machine learning in software testing,''
  in \emph{2008 The Eighth International Conference on Quality Software}, Aug
  2008, pp. 3--10.

\bibitem{Barr2015a}
E.~T. Barr, M.~Harman, P.~McMinn, M.~Shahbaz, and S.~Yoo, ``{The oracle problem
  in software testing: A survey},'' \emph{IEEE Transactions on Software
  Engineering}, vol.~41, no.~5, pp. 507--525, 2015.

\bibitem{automatedoracles}
M.~Pezzè and C.~Zhang, ``Automated test oracles: A survey,'' in \emph{Advances
  in Computers}.\hskip 1em plus 0.5em minus 0.4em\relax Elsevier, 2014,
  vol.~95, pp. 1 -- 48.

\bibitem{asan}
K.~Serebryany, D.~Bruening, A.~Potapenko, and D.~Vyukov, ``Addresssanitizer: A
  fast address sanity checker,'' in \emph{Proceedings of the 2012 USENIX
  Conference on Annual Technical Conference}, ser. USENIX ATC'12, 2012.

\bibitem{msan}
E.~Stepanov and K.~Serebryany, ``Memorysanitizer: fast detector of
  uninitialized memory use in c++,'' in \emph{Proceedings of the 2015 IEEE/ACM
  International Symposium on Code Generation and Optimization (CGO)}, San
  Francisco, CA, USA, 2015, pp. 46--55.

\bibitem{assertions}
D.~S. Rosenblum, ``A practical approach to programming with assertions,''
  \emph{IEEE Trans. Softw. Eng.}, vol.~21, no.~1, pp. 19--31, Jan. 1995.

\bibitem{bowring}
J.~F. Bowring, J.~M. Rehg, and M.~J. Harrold, ``Active learning for automatic
  classification of software behavior,'' in \emph{Proceedings of the 2004 ACM
  SIGSOFT International Symposium on Software Testing and Analysis}, ser. ISSTA
  '04, 2004, pp. 195--205.

\bibitem{Hu2008}
J.~Hu, W.~Yi, N.~W. Chen, Z.~J. Gou, and W.~Shuo, ``{Artificial neural network
  for automatic test oracles generation},'' \emph{Proceedings - International
  Conference on Computer Science and Software Engineering, CSSE 2008}, vol.~2,
  no.~05, pp. 727--730, 2008.

\bibitem{Braga2018}
R.~Braga, P.~S. Neto, R.~Rab{\^{e}}lo, J.~Santiago, and M.~Souza, ``{A machine
  learning approach to generate test oracles},'' \emph{ACM International
  Conference Proceeding Series}, pp. 142--151, 2018.

\bibitem{annoracle}
S.~R. Shahamiri, W.~M. N.~W. Kadir, S.~Ibrahim, and S.~Z.~M. Hashim, ``An
  automated framework for software test oracle,'' \emph{Information and
  Software Technology}, vol.~53, no.~7, pp. 774 -- 788, 2011.

\bibitem{daikon}
M.~D. Ernst, J.~H. Perkins, P.~J. Guo, S.~McCamant, C.~Pacheco, M.~S. Tschantz,
  and C.~Xiao, ``The daikon system for dynamic detection of likely
  invariants,'' \emph{Sci. Comput. Program.}, vol.~69, no. 1-3, pp. 35--45,
  Dec. 2007.

\bibitem{groce}
A.~{Groce}, T.~{Kulesza}, C.~{Zhang}, S.~{Shamasunder}, M.~{Burnett},
  W.~{Wong}, S.~{Stumpf}, S.~{Das}, A.~{Shinsel}, F.~{Bice}, and K.~{McIntosh},
  ``You are the only possible oracle: Effective test selection for end users of
  interactive machine learning systems,'' \emph{IEEE Transactions on Software
  Engineering}, vol.~40, no.~3, pp. 307--323, March 2014.

\bibitem{oraclenumber}
M.~{Harman}, S.~G. {Kim}, K.~{Lakhotia}, P.~{McMinn}, and S.~{Yoo},
  ``Optimizing for the number of tests generated in search based test data
  generation with an application to the oracle cost problem,'' in \emph{2010
  Third International Conference on Software Testing, Verification, and
  Validation Workshops}, April 2010, pp. 182--191.

\bibitem{readable}
S.~{Afshan}, P.~{McMinn}, and M.~{Stevenson}, ``Evolving readable string test
  inputs using a natural language model to reduce human oracle cost,'' in
  \emph{2013 IEEE Sixth International Conference on Software Testing,
  Verification and Validation}, March 2013, pp. 352--361.

\bibitem{McMinn2010}
P.~McMinn, M.~Stevenson, and M.~Harman, ``{Reducing qualitative human oracle
  costs associated with automatically generated test data},'' \emph{1st
  International Workshop on Software Test Output Validation, STOV 2010, in
  Conjunction with the 2010 International Conference on Software Testing and
  Analysis, ISSTA 2010}, pp. 1--4, 2010.

\bibitem{crowdoracles}
F.~Pastore, L.~Mariani, and G.~Fraser, ``Crowdoracles: Can the crowd solve the
  oracle problem?'' in \emph{Proceedings of the 2013 IEEE Sixth International
  Conference on Software Testing, Verification and Validation}, ser. ICST '13,
  2013, pp. 342--351.

\bibitem{Staats2012}
M.~Staats, G.~Gay, and M.~P. Heimdahl, ``{Automated oracle creation support,
  or: How I learned to stop worrying about fault propagation and love mutation
  testing},'' \emph{Proceedings - International Conference on Software
  Engineering}, pp. 870--880, 2012.

\bibitem{fbdebug}
A.~Roychoudhury and S.~Chandra, ``Formula-based software debugging,''
  \emph{Communications of the ACM}, vol.~59, no.~7, pp. 68--77, 2016.

\bibitem{darwin}
D.~Qu, A.~Roychoudhury, Z.~Lang, and K.~Vaswani, ``Darwin: An approach for
  debugging evolving programs,'' in \emph{Proceedings of the Symposium on
  Foundations of Software Engineering (ESEC/FSE)}.\hskip 1em plus 0.5em minus
  0.4em\relax Association for Computing Machinery, Inc., September 2009.

\bibitem{errorinvariants}
E.~Ermis, M.~Sch{\"a}f, and T.~Wies, ``Error invariants,'' in \emph{FM 2012:
  Formal Methods}, 2012, pp. 187--201.

\bibitem{angelic}
S.~Chandra, E.~Torlak, S.~Barman, and R.~Bodik, ``Angelic debugging,'' in
  \emph{Proceedings of the 33rd International Conference on Software
  Engineering}, ser. ICSE '11, 2011, pp. 121--130.

\bibitem{aflfast}
M.~B{\"o}hme, V.-T. Pham, and A.~Roychoudhury, ``Coverage-based greybox fuzzing
  as markov chain,'' in \emph{Proceedings of the 2016 ACM SIGSAC Conference on
  Computer and Communications Security}, ser. CCS '16, 2016, pp. 1032--1043.

\bibitem{aflgo}
M.~B{\"o}hme, V.-T. Pham, M.-D. Nguyen, and A.~Roychoudhury, ``Directed greybox
  fuzzing,'' in \emph{Proceedings of the 24th ACM Conference on Computer and
  Communications Security}, ser. CCS, 2017, pp. 1--16.

\bibitem{aflsmart}
V.-T. Pham, M.~B{\"o}hme, A.~E. Santosa, A.~R. C\u{a}ciulescu, and
  A.~Roychoudhury, ``Smart greybox fuzzing,'' \emph{IEEE Transactions on
  Software Engineering}, pp. 1--17, 2019.

\bibitem{bugex}
J.~R{\"o}{\ss}ler, G.~Fraser, A.~Zeller, and A.~Orso, ``Isolating failure
  causes through test case generation,'' in \emph{Proceedings of the 2012
  International Symposium on Software Testing and Analysis}, ser. ISSTA 2012,
  2012, pp. 309--319.

\bibitem{f3}
W.~Jin and A.~Orso, ``F3: Fault localization for field failures,'' in
  \emph{Proceedings of the 2013 International Symposium on Software Testing and
  Analysis}, ser. ISSTA 2013, 2013, pp. 213--223.

\bibitem{hercules}
V.-T. Pham, W.~B. Ng, K.~Rubinov, and A.~Roychoudhury, ``Hercules: Reproducing
  crashes in real-world application binaries,'' in \emph{Proceedings of the
  37th International Conference on Software Engineering - Volume 1}, ser. ICSE
  '15, 2015, pp. 891--901.

\bibitem{hampi}
A.~Kiezun, V.~Ganesh, P.~J. Guo, P.~Hooimeijer, and M.~D. Ernst, ``Hampi: A
  solver for string constraints,'' in \emph{Proceedings of the Eighteenth
  International Symposium on Software Testing and Analysis}, ser. ISSTA '09,
  2009, pp. 105--116.

\bibitem{boolector}
R.~Brummayer and A.~Biere, ``Boolector: An efficient smt solver for bit-vectors
  and arrays,'' in \emph{International Conference on Tools and Algorithms for
  the Construction and Analysis of Systems}.\hskip 1em plus 0.5em minus
  0.4em\relax Springer, 2009, pp. 174--177.

\bibitem{shalev}
S.~Shalev-Shwartz and S.~Ben-David, \emph{Understanding Machine Learning: From
  Theory to Algorithms}.\hskip 1em plus 0.5em minus 0.4em\relax New York, NY,
  USA: Cambridge University Press, 2014.

\bibitem{mcpa}
\BIBentryALTinterwordspacing
M.~B{\"o}hme, ``{MCPA}: Program analysis as machine learning,'' \emph{CoRR},
  vol. abs/11911.04687, 2019. [Online]. Available:
  \url{http://arxiv.org/abs/1911.04687}
\BIBentrySTDinterwordspacing

\bibitem{stads}
M.~B\"{o}hme, ``{STADS}: Software testing as species discovery,'' \emph{ACM
  Transactions on Software Engineering and Methodology}, vol.~27, no.~2, pp.
  7:1--7:52, Jun. 2018.

\bibitem{angluin}
D.~Angluin, ``Learning regular sets from queries and counterexamples,''
  \emph{Inf. Comput.}, vol.~75, no.~2, pp. 87--106, Nov. 1987.

\bibitem{abstention}
\BIBentryALTinterwordspacing
S.~Shekhar, M.~Ghavamzadeh, and T.~Javidi, ``Active learning for binary
  classification with abstention,'' \emph{CoRR}, vol. abs/1906.00303, 2019.
  [Online]. Available: \url{http://arxiv.org/abs/1906.00303}
\BIBentrySTDinterwordspacing

\end{thebibliography}

\end{document}